\documentclass[conference]{IEEEtran}
\IEEEoverridecommandlockouts
\usepackage{cite}
\usepackage{amsmath,amssymb,amsfonts}
\usepackage{algorithmic}
\usepackage{graphicx}
\usepackage{textcomp}
\usepackage{xcolor}
\usepackage{tikz}
 \usepackage{multirow}
 \usepackage{colortbl}
\usepackage[font=footnotesize]{caption}
\usepackage{caption}
\usepackage{subcaption}

\usepackage[labelsep=period]{caption}

\definecolor{color_29791}{rgb}{0,0,0}
\def\BibTeX{{\rm B\kern-.05em{\sc i\kern-.025em b}\kern-.08em
    T\kern-.1667em\lower.7ex\hbox{E}\kern-.125emX}}
\begin{document}

\title{Speaker Recognition in Realistic Scenario
Using Multimodal Data\\
}

\author{Saqlain Hussain Shah$^{1}$, Muhammad Saad Saeed$^{2}$, Shah Nawaz$^{3}$, Muhammad Haroon Yousaf$^{1}$\\
\{saqlain.hussain, saad.saeed, haroon.yousaf\}@uettaxila.edu.pk, shah.nawaz@desy.de
\\
$^{1}$University of Engineering and Technology Taxila, $^{2}$Swarm Robotics Lab NCRA,\\ 
$^{3}$Deutsches Elektronen-Synchrotron DESY
}



\IEEEoverridecommandlockouts
\IEEEpubid{\makebox[\columnwidth]{979-8-3503-2212-5/23/\$31.00~\copyright2023 IEEE \hfill} \hspace{\columnsep}\makebox[\columnwidth]{ }}

\maketitle

\IEEEpubidadjcol

\begin{abstract}
In recent years, an association is established between faces and voices of celebrities leveraging large scale audio-visual information from YouTube. The availability of large scale audio-visual datasets is instrumental in developing speaker recognition methods based on standard Convolutional Neural Networks. 
Thus, the aim of this paper is to leverage large scale audio-visual information to improve speaker recognition task.
To achieve this task, we proposed a two-branch network to learn joint representations of faces and voices in a multimodal system.
Afterwards, features are extracted from the two-branch network to train a classifier for speaker recognition. 
We evaluated our proposed framework on a large scale audio-visual dataset named VoxCeleb$1$.
Our results show that addition of facial information improved the performance of speaker recognition. Moreover, our results indicate that there is an overlap between face and voice.
\end{abstract}


\section{Introduction}
Speaker recognition is a fundamental task of speech processing with applications in a variety of real world domains. However,speaker recognition task is challenging under real world scenarios due to intrinsic and extrinsic variations. Intrinsic variations are associated with the speaker attributes namely gender, age and manner of speaking while extrinsic variations include factors outside the speaker personality such as background noise, microphone noise etc.~\cite{ngiam2011multimodal}. This makes speech signals prone to a large degree of variability. In recent years, Convolutional Neural Networks (CNNs) have opened new paths for speaker recognition task where speech signal is converted to spectrograms to be classified with these networks~\cite{nagrani2017voxceleb,chung2018voxceleb2}. 
Though, speaker recognition methods based on CNNs have surpassed the traditional methodologies~\cite{nagrani2017voxceleb}. However these methods suffered deterioration under real world scenarios. 
\begin{figure*}
\centering
\includegraphics[scale=0.9]{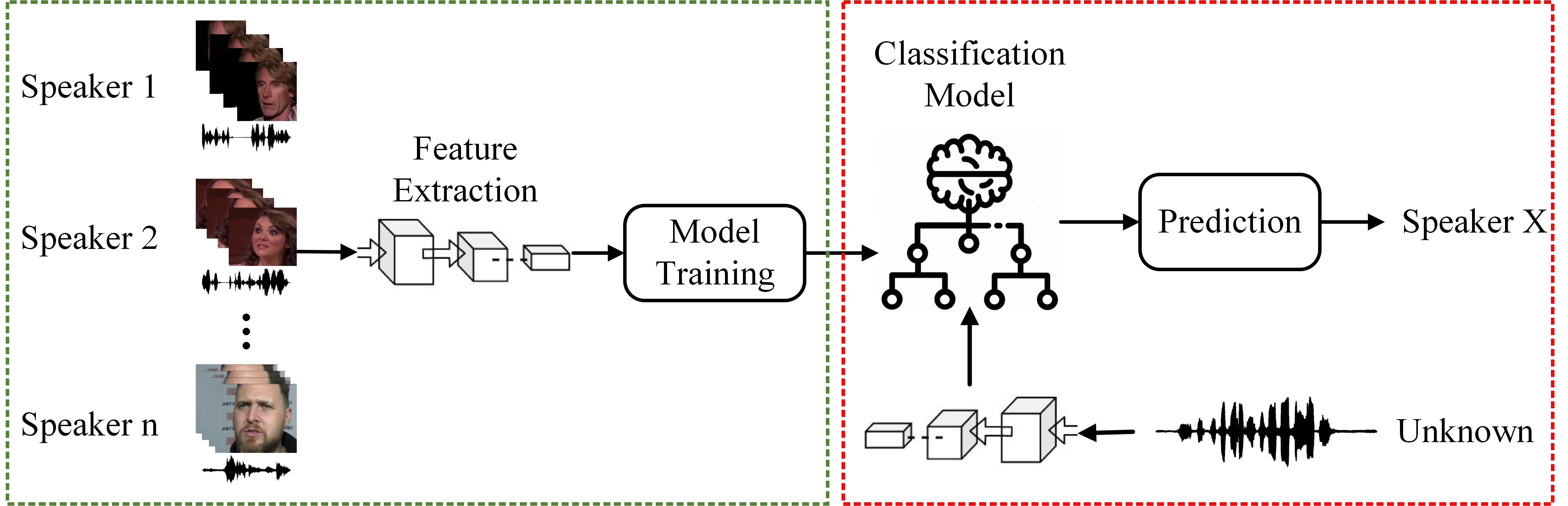}
\caption{The training and testing strategy for the proposed study. (Green) Shows the face tracks used for training the model. Both audio and visual modalities are used during training phase (Red) Only Audio Modality is available during phase. This protocol will help in knowing the impact of one modality, on performance of another modality.}
\label{fig:training_strat}
\end{figure*}
Recently, large scale datasets namely VoxCeleb$1$ and VoxCeleb$2$ are curated for speaker recognition task.  
These datasets are instrumental in developing CNN methods for speaker recognition task. 
For example, the work in~\cite{nagrani2017voxceleb,chung2018voxceleb2}  modified standard CNN such as VGG-M~\cite{chatfield2014return} and ResNet~\cite{he2016deep} to perform speaker recognition task. 
Moreover, both VoxCeleb$1$ and VoxCeleb$2$ datasets contain visual information which is instrumental for developing various multimodal applications such as cross modal transfer between face and voice~\cite{nawaz2019deep,nagrani2018seeing,wen2018disjoint,nawaz2021cross,nagrani2018learnable}, emotion recognition~\cite{albanie2018emotion}, speech separation~\cite{afouras2018conversation} and face generation~\cite{Wiles18a}.
These applications are instrumental in establishing a correlation between faces and voices of speakers.
Moreover, it is a well studied fact that humans end up associating voices and faces of people due to the fact that neuro-cognitive pathways for voices and faces share the same structure~\cite{kamachi2003putting}. 
Due to the availability of large scale audio-visual datasets such as VoxCeleb$1$ and association between faces and voices of speakers, a fundamental question arises: \textit{can audio-visual information can be used to improve speaker recognition task?}
To investigate it, we proposed a two-branch network to establish association between faces and voices. 
The proposed two-branch consists of the following three components: 1) feature extraction of faces and voices with task-specific pre-trained subnetworks, 2) a series of fully connected layers for faces and voices to learn joint multimodal representations, and 3) loss formulations. 
Afterwards, we extracted the features of audio segments to train a classifier for speaker identification task. Our results indicate that the facial information along with speech segments is instrumental in improve speaker recognition task. Fig.~\ref{fig:training_strat} shows the training and testing strategy of the proposed framework.

We summarize our key contributions as follows: 1) We propose a two-branch network to learn multimodal discriminative joint representations of faces and voices of speakers. 2) We present a comparisons of speaker recognition task with only speech segments and multimodal information. 3) Our results indicate that multimodal information considerable improves speaker recognition task.  

The paper is organized in the following sections. Section~\ref{sec:related_work} provides detail overview of the the related work. Section~\ref{sec:framework} provides an overview of the proposed framework following by results and discussion in Section~\ref{sec:experuments}. Finally, Section~\ref{sec:conclusion} provides concluding remarks of our work.

\section{Related Work}
\label{sec:related_work}
We summarize previous work relevant to the speaker recognition and face-voice association tasks.
\subsection{Speaker Recongition}
Sandra et al.~\cite{pruzansky1963pattern} laid the groundwork for speaker recognition systems attempting to find a similarity measure between two speech signals by using filter banks and digital spectrograms. We provided a brief overview of speaker recognition methods as clustered in two main categorizes: Traditional and neural network based methods.\\
\noindent \textbf{Traditional methods.~}There have been many advancements in the speaker recognition task due to the availability of data and computing resources. However, noisy environment presents a challenging scenario. For several years, the standard speaker recognition task relied on ways that are dependant on features that required manual intervention and domain knowledge. This includes features extracted from low dimensional short term representation of the speech signals such as MEL Frequency Cepstrum Coefficients~\cite{mammone1996robust}. 
The performance of these systems
degrade in real world conditions~\cite{yapanel2002high,hansen2001robust}. 
These systems are dependent on Human ability to extract useful features which is a limitation for the system.  
Joint Factor Analysis captures both speaker-specific and session-specific variability in speech signals by decomposing the speech signal into a set of latent factors~\cite{kenny2005joint}. 
Support Vector Machine (SVM) classifier has been very successful for robust recognition tasks.  However, such methods are very slow, complex and prone to degradation when applied to various real world scenarios. 
Despite of these advancements, the performance of traditional approaches drop in the presence of noise. Moreover, the performance degraded as the size of data increases. In real world applications there is often no knowledge of environmental noise, transmission channel used and number of speakers in the background. In such cases the traditional methods may degrade in performance. \\
\noindent \textbf{Deep Learning Methods.~}Over the last few years, advances in computing resources and neural networks has led to  more efficient methods. With these advancement, CNN are extensively used in tasks such as speaker recognition. 
For example, the work in~\cite{nagrani2017voxceleb,chung2018voxceleb2} propose a CNN based method to transform speech segments into spectrograms for speaker recognition task. 
With this advancement, speaker recognition task is moved from manually extracted features to data driven methods. 
Specifically, the work in~\cite{nagrani2017voxceleb} train a modified VGG-M on spectrogram extracted directly from speech segment. 

\subsection{Face-voice Association}
Recently, an association between faces and voices of speakers are established by leveraging cross-modal verification and matching tasks~\cite{nagrani2018seeing,nagrani2018learnable,nawaz2019deep,nawaz2021cross,saeed2022fusion,saeed2022learning}. 
The work in~\cite{nagrani2018seeing} used a triplet network to learn joint representation for face-voice association task. 
Similarly, the work in~\cite{kim2018learning} used a triplet network~\cite{hoffer2015deep} to minimize the distance between faces and voices by extracting features from face subnetwork \cite{simonyan2014very} and voice subnetwork~\cite{aytar2016soundnet}. 
Nawaz et. al~\cite{nawaz2019deep} learns shared latent space by taking advantage from class centers with a single stream network which eliminate the need of pair or triplet samples.
On similar grounds, Saeed et. al~\cite{saeed2022fusion,saeed2022learning} proposed a light-weight, plug-and-play mechanism that exploits the complementary cues from faces and voices
to form enriched fused embeddings and clusters them based on their identity labels via orthogonality constraints.

In contrast to existing methods, our goal is to extract robust features from a multimodal system trained on faces and voices for speaker recognition task.

\begin{figure*}
     \centering
     \begin{subfigure}[b]{0.45\textwidth}
         \centering
         \includegraphics[width=\textwidth]{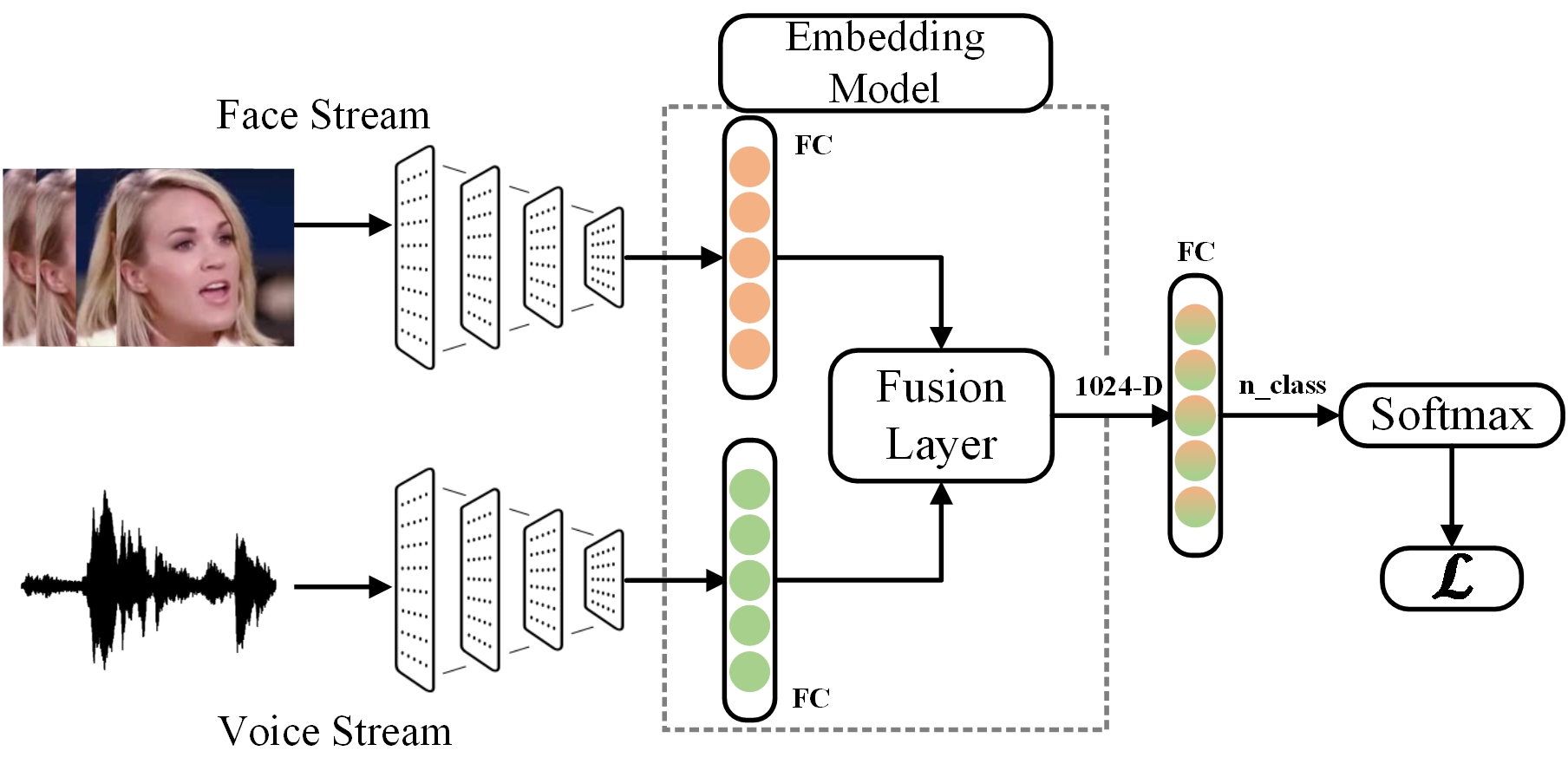}
         \caption{Proposed Two-branch network}
         \label{fig:two-branch}
     \end{subfigure}
     \begin{subfigure}[b]{0.45\textwidth}
         \centering
         \includegraphics[width=\textwidth]{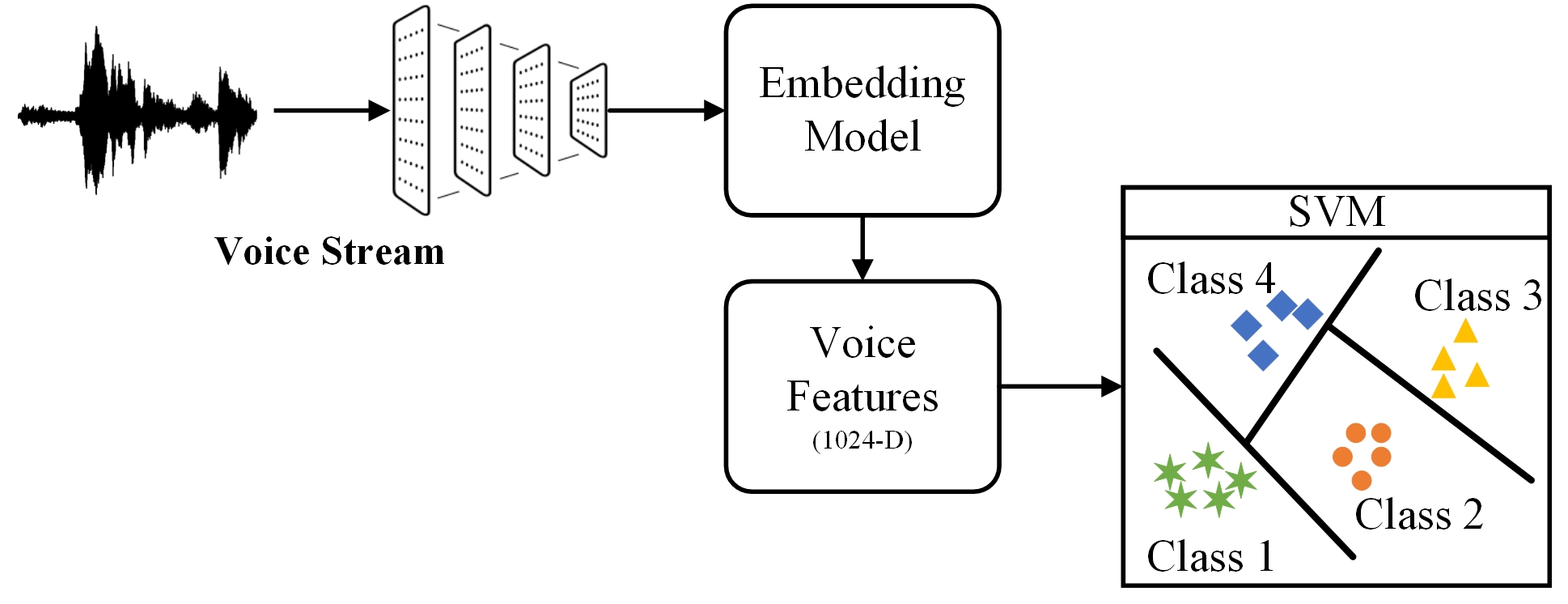}
         \caption{Testing strategy with single modality}
         \label{fig:single-branch}
     \end{subfigure}
        \caption{
        (a) Independent modality-specific embedding networks are leverage for off-the-shelf feature extraction (Box) The proposed \textbf{Two-branch Model} with independent modality-specific FC layers. \textbf{Element-wise multiplication} is used for Fusion of two branches.
        (b) During testing phase only audio data is used. The visual data is set to 0. Features of audio samples from training and testing splits are extracted. Later on, an SVM is trained on these features to report \textbf{{\%}} accuracy.
        }
        \label{fig:main_arch}
    
\end{figure*}


\section{Overall Framework}
\label{sec:framework}
\subsection{Baseline}
We extracted $1024$-D features of VoxCeleb$1$ dataset with VGGVox subnetwork to establish a baseline. SVM classifier is trained on these features for speaker recognition task. Decision function shape is set to one vs one for multi class classification using SVM, kernel parameter was set to poly while degree of polynomial kernel function was set to 3. After training SVM on the features extracted using VGGVox, accuracy of identification is 91\%.
\subsection{Two Branch Network}
Our proposed method consists of training a multimodal system using a two-branch network with face and voice information.
Afterwards, the multimodal is used to extract features to train a classifier for speaker recognition task.  
Face and audio features are extracted from VGGFace~\cite{parkhi2015deep} and VGGVox~\cite{nagrani2017voxceleb} subnetworks respectively. 
Afterwards, face and voice features are input to two independent branches, with each modality specific branch respectively. 
Features from both subnetworks are fused after passing from fully connected and normalization layers. Fig.\ref{fig:main_arch} shows the proposed framework. 
\subsection{Multimodal Fusion}
We extracted features from face and voice information. These features are then fused and passed to fully connected layer to learn joint representations from both face and voices signals. After fusion a softmax layer is used to learn for the output classes. 
Softmax function is used as the activation function to predict a multinomial probability distribution where probability is required for multi class classification problems. Features extracted from this two branch network are then used to train a classifier. 
\subsection{Loss Formulation}
We want the fused feature to capture the semantics of the speaker or identity. In other words, these features should be able to predict the identity labels with good accuracy. It is possible if the samples belonging to the same class are placed nearby whereas the ones with different classes are far away. A popular choice to achieve this is softmax cross entropy (CE) loss, which also allows stable and efficient training. Now, the
loss with fused embeddings is computed as
\begin{equation}
  \mathcal{L}_{CE}  =  \sum_{i}^C \mathbf{l}_{i} log(f(\mathbf{l}_{i}) ,
    \label{Eq:OC}
\end{equation}
Categorical cross entropy is a very good measure of how distinguishable two discrete probability distributions are from each other~\cite{zhang2018generalized}. Adam was used as optimizer with learning rate ranging from $0.01$ to $0.13$. 
Network was trained using batch size of $512$, $1024$, $2048$ and $4096$. 
Maximum results were achieved with $0.04$ as learning rate and $2048$ as batch size.

\section{Experiments and Results Discussions}
\label{sec:experuments}
\subsection{Training Detail and Dataset}
\noindent \textbf{Dataset.}
VoxCeleb$1$ is a large-scale dataset of audio-visual human speech videos extracted ‘in the wild’ from YouTube. 
These videos contain real world noise with background chatter, overlapping speech, laughter and recording equipment noise. Table~\ref{tab:voxceleb1} provides shows statistics of the dataset.

\begin{table}[!t]
\caption{VoxCeleb$1$ Identification Split}
\centering
\begin{tabular}{|c|c|c|c|}
\hline
 & \textbf{Train} & \textbf{Test} & \textbf{Total}\\
 \hline
 \# of speakers & 1,251 & 1,251 & 1,251\\
 \hline
 \# of videos & 21,245 & 1,251 & 22,496\\
 \hline
 \# of utterances & 145,265 & 8,251 & 153,516\\
 \hline
\end{tabular}
\label{tab:voxceleb1}
\end{table}

\begin{figure*}
     \centering
     \begin{subfigure}[b]{0.45\textwidth}
         \centering
         \includegraphics[width=\textwidth]{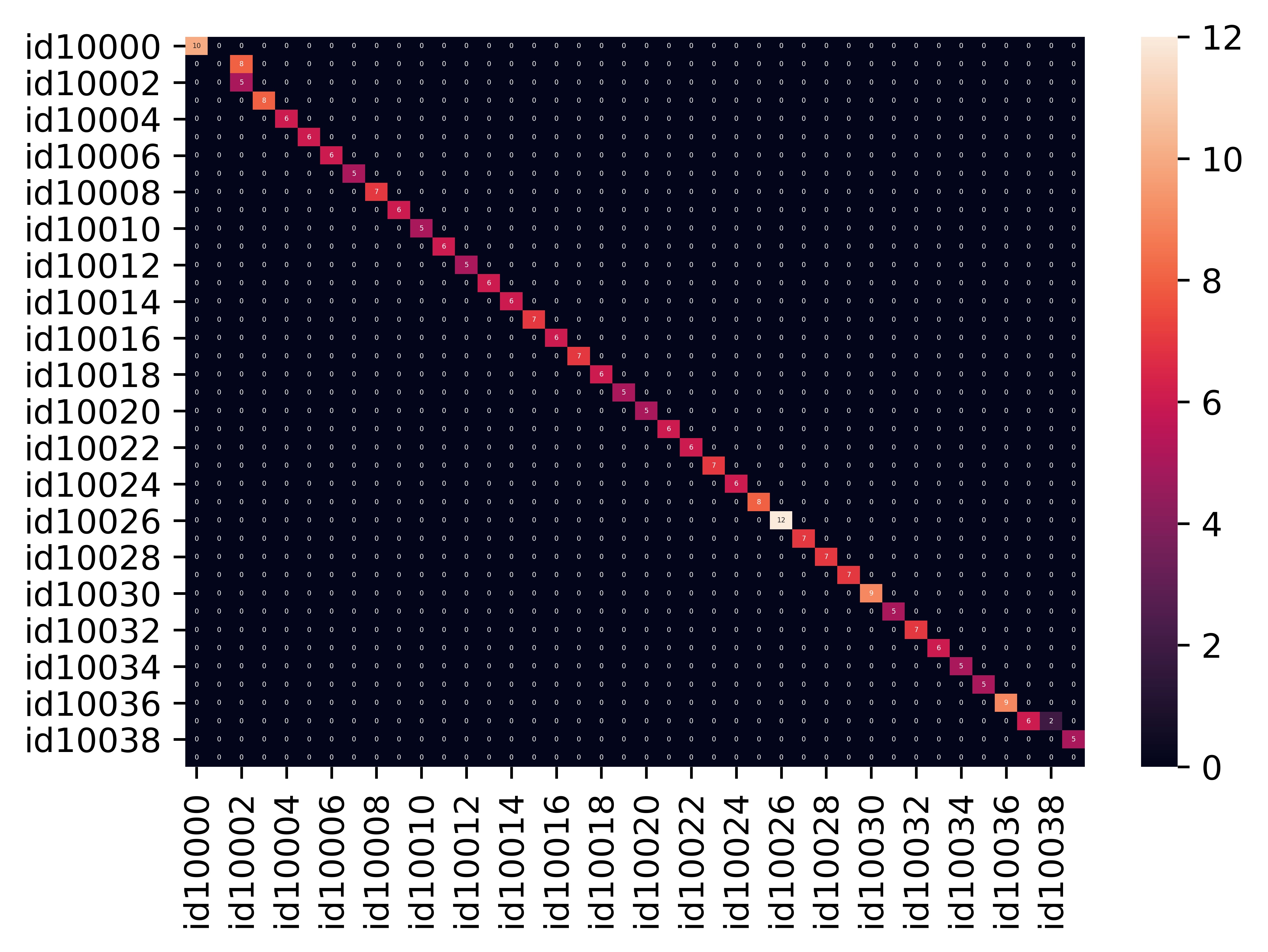}
         \caption{Confusion matrix of voice segments}
         \label{subfig:confusin_baseline}
     \end{subfigure}
     \begin{subfigure}[b]{0.45\textwidth}
         \centering
         \includegraphics[width=\textwidth]{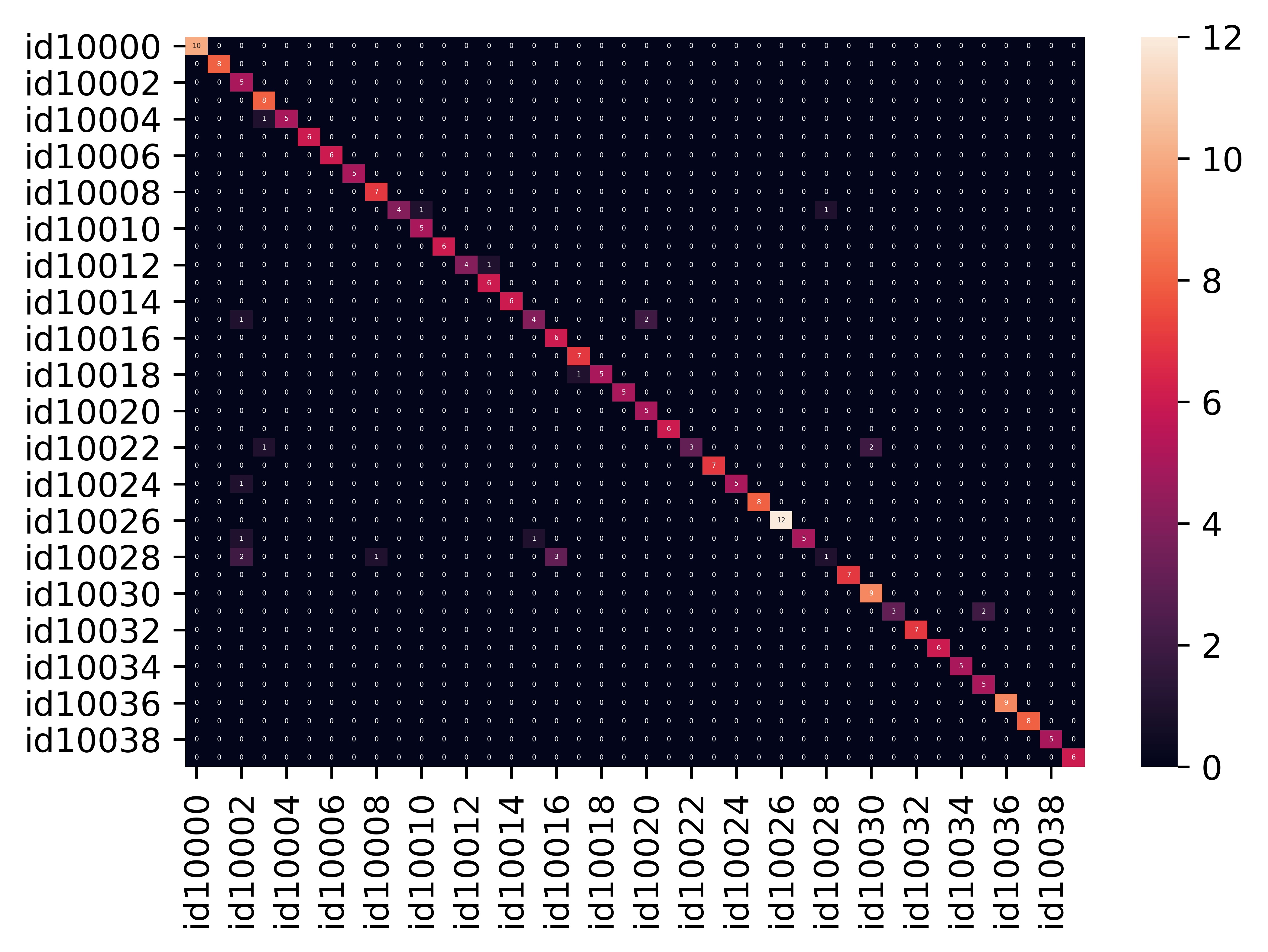}
         \caption{Confusion matrix of voice segment extracted from two-branch network}
         \label{subfig:confusion_multimodal}
     \end{subfigure}

\caption{
        Confusion Matrix of test data: (left) Confusion Matrix of features extracted from VGGVox Network (right) Confusion Matrix of features extracted from proposed two-branch network trained on multimodal data. Confusion matrix shows results for $20$ identities. (Best viewed in color and zoomed in)
        }
     
\end{figure*}
\begin{figure*}[!t]
     \centering
     \begin{subfigure}[b]{0.4\textwidth}
         \centering
         \includegraphics[width=\textwidth]{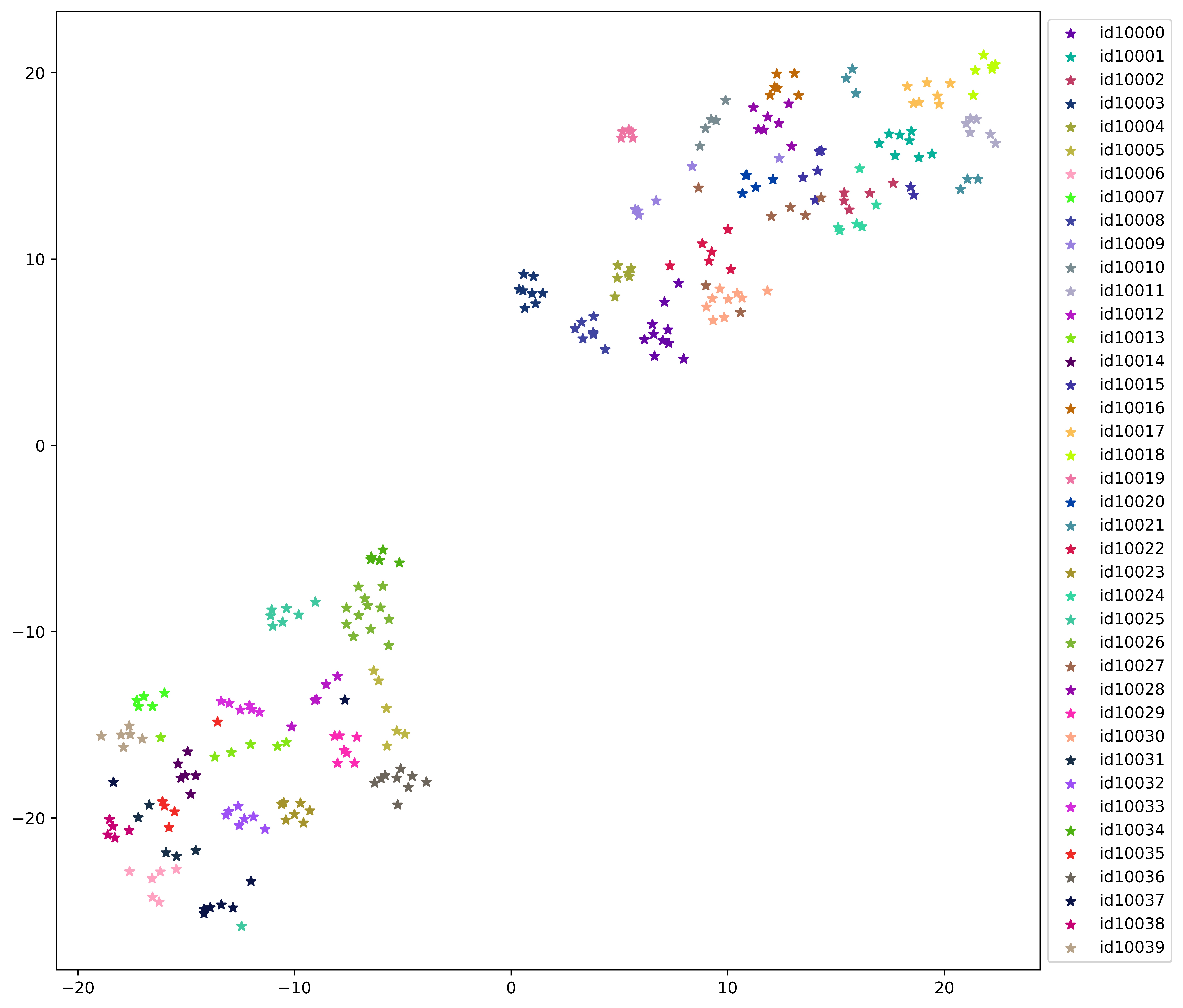}
        \caption{Visualization of of voice segments}

         \label{fig:tnse_voice}
     \end{subfigure}
     \begin{subfigure}[b]{0.4\textwidth}
         \centering
         \includegraphics[width=\textwidth]{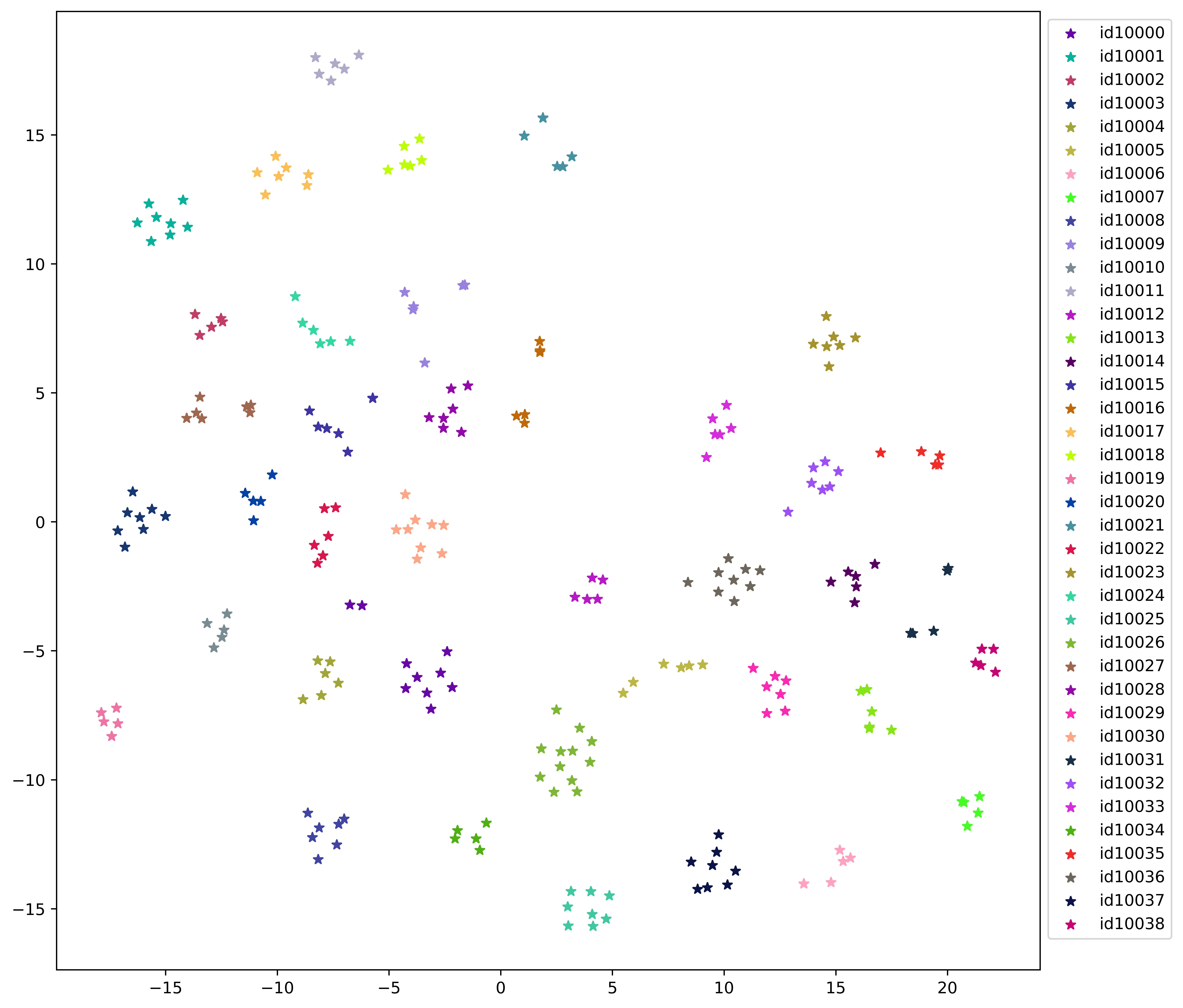}
         \caption{Visualization of voice segment extracted from two-branch network}

         \label{fig:tnse_multimodal}
     \end{subfigure}
        \caption{
        T-SNE plot of test data: (left) T-SNE plot of voice features extracted from VGGVox Network (right) T-SNE plot of voice features extracted from proposed two-branch network trained on multimodal data. Features of $30$ random identities are selected.
        }
        \label{fig:tnse}  
\end{figure*}

\noindent  \textbf{Training.} Inspired by~\cite{saeed2022fusion}, we propose a two-branch network to analyze the effect of multimodal information on speaker recognition task. 
Face embeddings were extracted from pretrained VGGFace~\cite{parkhi2015deep} while audio embeddings are extracted from VGGVox~\cite{nagrani2017voxceleb}. 
Face and voice embeddings were passed as input to two-branch network with subnetworks containing multiple dense layers followed by dropout and normalization layers. Dropout of 10\% and 20\% was used during training. 
Normalized embeddings from two subnetworks are then fused and passed through dense and normalization layers to softmax layer containing $1251$ hidden units for the classes in dataset. Training is performed with multiple margin values, dropout, batch size, loss functions and learning rates. \\
\noindent \textbf{Testing}
Features are extracted from the two branch network to train and test support vector machine classifier. We extracted feature with $1024$-D size form the fusion layer of the model. 
Feature extraction is performed in two ways:
\begin{itemize}	
\item \emph{Aiding with face signals:} During this phase face and speech signals were provided as input to trained two branch network and speech features were extracted from it.
\item \emph{Aiding without face signals:} During this phase only speech signals were provided as input to trained two branch network and speech features were extracted from it. For face subnetwork input vector was set to zero.
\end{itemize}

\noindent \textbf{Speaker identification.} Features extracted are normalized and used to train and test a SVM classifier. 
Kernel parameter of support vector machine is set to \emph{poly} while decision function shape was set to \emph{ovo}. Remaining parameters are set to default during training. 

\subsection{Results from Voice Only Features}
We extracted the feature from VGGVox subsnetwork and trained a classifier to establish a baseline, resulting in 91\% identification performance. Fig.~\ref{subfig:confusin_baseline} shows confusion matrix of the baseline results.
Moreover, stochastic neighbor embedding for a sample test set in Fig.~\ref{fig:tnse_voice} shows that the network has distributed the features for multiple classes far apart which has reduced the accuracy of classifier on those features.


\subsection{Results from Aided Facial Information}
Table~\ref{tab:main_results} shows results of a classifer trained on features extracted from the two-branch network. 
Confusion matrix of two branch fused features can be seen in Fig.\ref{subfig:confusion_multimodal}. Moreover, T-SNE plot for a sample test set in Fig.\ref{fig:tnse_multimodal} shows that the network has distributed the features more efficiently where same class features are close to each other that has resulted in better learning of SVM for speaker recognition task.

Experiments show that when speech signals are aided by faces during feature extraction, speaker recognition is improved significantly. Without facial information, the system is likely to be effected with noise. When we have face information aiding the voice information degraded in one mode can be recovered by the other.

\begin{table}[!t]
\caption{Speaker identification performance on VoxCeleb$1$. (Higher is better)}
\centering
\begin{tabular}{|c|c|c|}
\hline
\textbf{Method} & \textbf{Loss} & \textbf{Top-1} \% \\
 \hline
 I-vectors + PLDA + SVM~\cite{nagrani2017voxceleb} & - & 60.8\\
 \hline
 CNN~\cite{nagrani2017voxceleb} & - & 80.5 \\
 \hline
 VGGVox(Baseline) & - & 91.0 \\
 \hline
 Network A (ndims=128)~\cite{yadav2018learning} & Center+Softmax& 84.6 \\
 \hline
 Network B (ndims=128)~\cite{yadav2018learning} & Center+Softmax& 89.5 \\ 
 \hline
 {\textbf{Ours}} 
  & \textbf{CE Loss} & \textbf{97.2}\\
 \hline
\end{tabular}
\label{tab:main_results}
\end{table}

\section{Conclusion}
\label{sec:conclusion}
In this work we proposed that presence of multimodal information improve the performance of speaker recognition task. We propose the two-branch network to extract features from both face and voice signals. SVM was used to classify speaker based on features from single domain and multi domain. We obtained promising results when we used both face and speech information as input to our model. The identification performance achieved using our approach is higher compared to VGGVox which only exploit single modality. Also the results using both speech and face signals while extracting features from our model are better as compared to inputting only speaker information which clearly indicates that face information can aid in speaker recognition. Also, this increase in speaker recognition performance with the aid of facial information gives us clue that though there is some association between face and voice of a person.
Another very important contribution is that this work opens the research path for classification and retrieval tasks of other modalities.\\
\textbf{Acknowledgements.}~
Authors gratefully acknowledge the support of Swarm Robotics Lab, NCRA for providing the  necessary equipment and resources for our experiments. 

\bibliographystyle{IEEEbib}
\bibliography{IEEEbib}
\end{document}